\documentclass[pre,a4paper,twocolumn,showkeys]{revtex4}
\usepackage{graphics}

\begin{document}

\title{Nature of Chemisorption on Titanium Carbide and Nitride}

\author{Carlo Ruberto}
\email{ruberto@fy.chalmers.se}
\author{Aleksandra Vojvodic}
\author{Bengt I. Lundqvist} 

\affiliation{Department of Applied Physics, Chalmers University of Technology, 
SE-412 96 G\"{o}teborg, Sweden}

\keywords{Density-functional calculations, Titanium carbide, 
Titanium nitride, Adatoms, Adsorption, Chemisorption, Density of states.}

\begin{abstract}

Extensive density-functional calculations are performed to understand 
atomic chemisorption on the TiC($111$) and TiN($111$) surfaces, in 
particular the calculated pyramid-shaped trends in the adsorption 
energies for second- and third-period adatoms.  
Our previously proposed concerted-coupling model for chemisorption on 
TiC($111$) is tested against new results for adsorption on TiN($111$) and found 
to apply on this surface as well, thus reflecting both similarities and 
differences in electronic structure between the two compounds.  

\end{abstract}

\maketitle

\section{Introduction}

Within an extensive theoretical study of atomic chemisorption on the 
Ti-terminated polar TiX($111$) (X = C, N) surfaces, 
we have presented characteristic trends in adsorption energies and geometries 
for H, second-row elements B, C, N, O, and F, and third-row elements 
Al, Si, P, S, Cl \cite{Ruberto,Vojvodic}.  The study has both fundamental 
and technological motivations, as these materials are important in 
applications in which surface properties are essential, 
such as growth of alumina on Ti(C,N)($111$) and use of TiN as a thin 
diffusion barrier in integrated circuits or as a biocompatible 
material.  

Our results show similar trends for TiC and TiN, with 
atomic adsorption energies for the second- and third-period elements 
varying along the periods in a characteristic pyramid-like way [Fig.\ 
\ref{fig:DeltaLDOS_PER2_TiN}(a)] \cite{Ruberto,Vojvodic}.  
Within each period, the energies increase from group III to group VI and 
decrease from group VI to group VII, with overall higher values 
for period 2.  These variations are very large, from $\sim 3.5$ eV for Al 
to $\sim 8.5$ eV for O.  
Also, the preference for adsorption on TiC {\it vs.}\ TiN varies, 
with a slightly stronger chemisorption on TiN for the elements 
toward the left of the periodic table (B and Al) and a slightly 
stronger chemisorption on TiC for the remaining adatoms.  

Such variations cannot be explained within a simple model, where only
one type of coupling mechanism is considered.  In a separate study,
we propose a concerted-coupling model for adsorption on TiC($111$),
in which two different types of adatom--substrate couplings are
active \cite{Ruberto}.  This is similar to the $d$-band model for
chemisorption on transition metals \cite{Hammer}, however, with
very different couplings, due to the more complex electronic structure
of TiC.
In this contribution, we pursue our investigation further, 
analyzing the electronic structure of the different adsorbates on TiN($111$) 
and discussing whether our proposed model can be generalized 
and used to understand the similarities and differences 
between the TiN($111$) and TiC($111$) substrates.

\section{Computational Method}

The computational method used is described in Ref.\ \cite{Vojvodic}.  
In short, the pseudopotential-plane-wave density-functional-theory (DFT) 
code {\tt dacapo} is used, with PW91 GGA exchange-correlation potential, 
ultrasoft pseudopotentials, and slab geometry \cite{dacapo}.  

In the present paper, the electronic structures of the considered systems 
are analyzed by calculating and plotting the total and local densities of states 
(DOS and LDOS, respectively) obtained from the calculated Kohn-Sham (KS) wavefunctions 
and energy eigenvalues.  Atom- and/or orbital-projected 
LDOS($E$)'s are obtained by projecting the KS wavefunctions onto the 
individual atomic orbitals and then 
plotting them as functions of the energy 
relative to the Fermi level, $E - E_F$.  The spatial extent of 
the DOS's at specific energy intervals is also analyzed in detail by examining 
the KS-wavefunction amplitudes 
for the relevant electronic states in three-dimensional 
real space.  Also, the charge localization around individual atoms 
is analyzed with the ``atoms-in-molecule'' method of Bader \cite{Bader}.

\section{Results and Analysis}

\subsection{Electronic structure of the clean TiX(111) surfaces}

Bulk TiC and TiN adopt the NaCl structure.  Thus, in the ($111$) direction, 
the crystals are composed of alternating Ti and X atomic layers, in an $ABC$ 
stacking sequence.  Our study focuses on the Ti-terminated surfaces, since 
experiments show this to be the stable termination for TiC($111$) \cite{Oshima}.  
The calculated lattice parameters and surface relaxations 
are given in Refs.\ \cite{Ruberto,Vojvodic}.  

\begin{figure*}
\scalebox{0.74}{\includegraphics{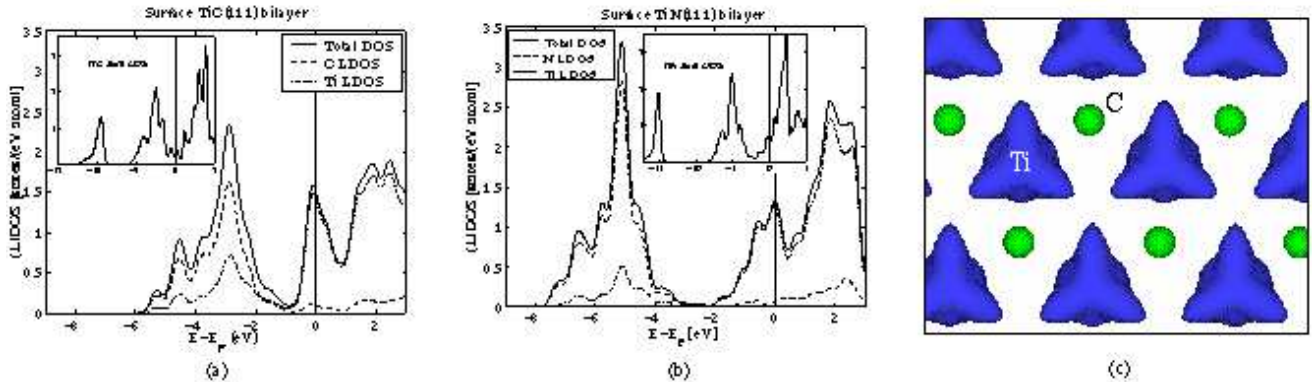}}
\caption{\label{fig:LDOS_surf}(a--b) Total and atom-projected DOS for the top TiX bilayer 
of the clean (a) TiC($111$) and (b) TiN($111$) surfaces. 
The inserts in (a) and (b) show the total DOS for bulk TiC and TiN, respectively.
(c) Real-space contour plot (from above the surface and showing 
only the top TiC bilayer) of the total charge density of the electronic 
states above $-0.6$ eV on TiC($111$), showing the TiSR (C atoms: green balls; 
Ti atoms: inside the electron clouds, as marked in the figure).}
\end{figure*}

Both Ti-terminated TiX($111$) surfaces are characterized by the presence of strong 
Ti-localized surface resonances (TiSR's) \cite{footnote_TiSR}, 
pinned at $E_F$ [Figs.\ \ref{fig:LDOS_surf}(a--b)] and 
consisting of mainly $d_{(xz,yz)}$ and $d_{(xy, x^2-y^2)}$ orbitals (where $z$ is 
perpendicular to the surface).  
Figure \ref{fig:LDOS_surf}(c) shows this symmetry and how the TiSR electron distribution is 
localized mainly around the Ti atoms (as evidenced also by the small 
width of the TiSR peaks) and extends toward the fcc surface sites.

Our calculated surface DOS's [Figs.\ \ref{fig:LDOS_surf}(a--b)] show 
that the TiSR is more filled on TiN($111$) than on TiC($111$).  
This is confirmed by a Bader analysis of the surface charges, yielding a 
surface-Ti ionicity of 
$+1.09$ for TiC and of $+0.96$ for TiN.  This can be understood by considering the 
filling of the TiSR's as the mechanism that compensates the electrostatic instability of the 
polar TiX($111$) surfaces \cite{Tsukada}.  
As bulk TiN is more ionic than bulk TiC 
(Ti $\rightarrow$ X charge transfers of $1.51$ electrons for TiC and $1.62$ for TiN, 
according to our Bader analysis)
\cite{Vojvodic}, a larger compensating surface charge is needed on TiN($111$).  

The bulk electronic structure of TiX 
[inserts in Figs.\ \ref{fig:LDOS_surf}(a--b)]
consists of 
(i) a lower valence band (LVB), of only X$2s$ character, 
(ii) an upper valence band (UVB), characterized by bonding Ti$3d$--X$2p$ states in its 
high-energy part and by X$2p$--X$2p$ bonding interactions in its low-energy part, and 
(iii) a conduction band (CB), of mainly Ti$3d$ character, corresponding to 
antibonding Ti$3d$--X$2p$ states \cite{Ruberto,Vojvodic}.  The ionicity of 
bulk TiX is evidenced by 
a predominance of Ti (X) character in the CB (UVB).  

\begin{figure*}
\scalebox{1.17}{\includegraphics{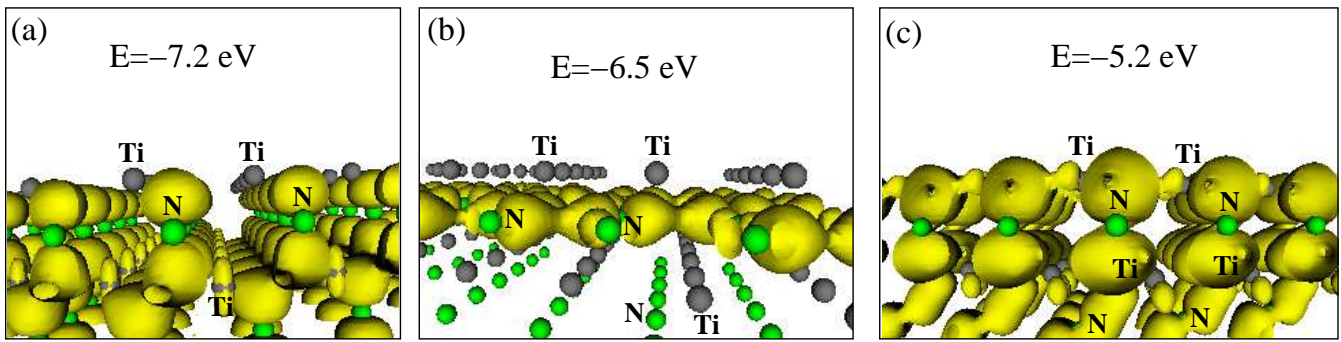}}
\caption{\label{fig:3DPlots}Real-space contour plots of the KS wavefunctions for the 
clean TiN($111$) surface at selected energies, viewed perpendicularly to the 
surface, illustrating:  
(a--b) typical bonding N--N interactions, with and without coupling to bulk states, 
in the lower UVB region;  
(c) a typical Ti--N bonding state, with high N-localization, in the main UVB peak.  
Ti (N) atoms are grey (green) balls, or lie inside the electron clouds, as marked in 
the figures.}
\end{figure*}

Both TiX($111$)-surface UVB's [Figs.\ \ref{fig:LDOS_surf}(a--b)] 
are composed of one main peak, three 
smaller peaks at lower energies, a shoulder on the high-energy side of the main peak, 
and a small peak at the upper edge of the UVB.  
For both surfaces, our detailed analysis of the individual KS wavefunctions composing the UVB 
(Fig.\ \ref{fig:3DPlots}) shows that, similarly to the bulk UVB, the main UVB peak and 
the high-energy UVB region consist of Ti--X bonding states [Fig.\ \ref{fig:3DPlots}(c)].  
On the other hand, 
the three low-energy peaks consist almost exclusively of X--X bonding states that are mainly 
localized to the surface region [Figs.\ \ref{fig:3DPlots}(a--b)], that is, they are X-localized 
surface resonances (XSR's).

Like for the bulk DOS \cite{Vojvodic}, several differences exist between the 
TiN($111$) and TiC($111$) surface DOS's [Figs.\ \ref{fig:LDOS_surf}(a--b)]:  
(i) the UVB and CB of TiN($111$) lie at lower energies, due to the partial filling of the CB 
with the extra electron per formula unit present in TiN; 
(ii) the pseudogap between the UVB and the TiSR is larger for TiN($111$);  and 
(iii) the UVB DOS is considerably larger and more X-localized on TiN, which again is an 
indication of the higher ionicity of TiN.  

Thus, both surface DOS's differ from their corresponding bulk DOS's [see Figs.\ 
\ref{fig:LDOS_surf}(a--b)] by (i) the presence of TiSR's around $E_F$, 
(ii) a quenching of the UVB peak lying above the main peak, and 
(iii) the presence of XSR's in the low-energy UVB region.  
These changes can be understood in terms of breakage of the Ti--X and X--X bonds that 
cross the ($111$) plane upon cleavage of the bulk structures to create the ($111$) surfaces.  
As a consequence, the bonding and antibonding Ti--X states, which are located in the 
high-energy part of the UVB and in the CB, respectively, collapse into more atomic-like 
X and Ti orbitals.  Therefore, the quenching of the high-energy UVB peak 
can be associated with the disappearance of X-localized bonding Ti--X states, while 
the appearance of the TiSR's arises from the energetical lowering of Ti-localized 
antibonding Ti--X states (indeed, for both compounds, a closer inspection of the CB 
reveals the quenching of a bulk CB peak located right above the TiSR's).  
The TiSR's can thus be interpreted as dangling bonds extending toward the fcc sites 
(which in the bulk would be occupied by X atoms) [Fig.\ \ref{fig:LDOS_surf}(c)].  
Similarly, the appearance of the XSR's in the low-energy UVB region is 
caused by the breakage of X--X bonding interactions, which dominate at 
these energies.

\subsection{Electronic structure of fcc adatoms on TiX(111)}

In a separate study, we have proposed a model for the description of the exceptionally 
strong atomic chemisorption on TiC($111$) \cite{Ruberto}.  
In this concerted-coupling 
model, based on the Newns-Anderson (NA) model \cite{Newns}, the adsorption is 
understood to arise from the concerted action of two different types of 
adatom--substrate coupling:  a strong adatom--TiSR coupling and weaker adatom--CSR 
couplings.  

Briefly, the NA model for chemisorption on metal surfaces yields two limiting cases: 
weak and strong chemisorption, corresponding, respectively, to small and large coupling 
matrix elements between adatom and substrate states, compared to the substrate band 
width.  In the weak chemisorption, the adsorbate level is shifted and broadened (if it 
lies in the energetical region of the substrate band).  In the strong chemisorption, 
the adatom level is replaced by two levels that lie on each side of the substrate 
band.  These are usually interpreted as the bonding and antibonding states of the 
``surface molecule'' formed by the adatom and its neighboring substrate atoms.  
These bonding and antibonding states appear gradually, as the adatom--substrate 
hopping integral $h$ increases.  At small $h$, the adlevel is broadened.  
As $h$ increases, this level broadens further and splits gradually into two 
separate levels \cite{Spanj_Desj}.  

Our concerted-coupling model for TiC($111$) is based on an analysis of the 
difference in DOS ($\Delta$DOS) before and after adsorption 
(thus, the $\Delta$DOS shows the appearance and quenching of states 
due to the adsorption of an adatom as positive and negative peaks, respectively) 
\cite{Ruberto}.  For all considered adatoms on TiC($111$), our $\Delta$DOS's show:  
(i) the appearance of Ti-dominated states at $\sim +1.0$ eV; 
(ii) a strong quenching of the TiSR; 
(iii) the appearance of states below $E_F$, whose energies 
decrease with increasing adatom number $Z$ within each adatom period 
(due to the decreasing free-atom energy as $Z$ increases).  

Within the NA model, these results are interpreted as evidence for a strong 
adatom--TiSR coupling, giving rise to antibonding and bonding states above 
and below, respectively, $E_F$.  At the same time, our calculated $\Delta$DOS's show 
also that the bonding adatom--TiSR 
level below $E_F$ is successively broadened and split into subpeaks as its 
energy approaches the middle of the substrate UVB region.  This indicates, 
within the NA model, an additional, weaker, coupling between the TiSR-modified 
adlevel and the substrate UVB peaks, which are characterized by CSR's.  
Thus, the subpeaks correspond to combinations of the bonding and 
antibonding solutions that arise from adatom--CSR couplings.  This is 
confirmed by (i) the presence, in the substrate-projected $\Delta$DOS's, 
of negative peaks that are almost exclusively localized around the C atoms and 
that lie at energies in between the subpeaks and (ii) our state-resolved, real-space, 
analysis of the DOS, which shows that the subpeaks are dominated by adatom--C 
bonding states at lower UVB energies and by adatom--Ti bonding states at higher 
energies.  The adatom--C states are only present at energies lower than the main 
substrate UVB peak and can only be interpreted as bonding solutions of adatom--CSR 
couplings.  This analysis shows also that the adatom--C bonding states get successively 
stronger as the energetical overlap between the TiSR-modified adlevel and the substrate 
CSR's increases.  Maximum overlap is achieved for the O adatom, for which all 
bonding adatom--TiSR states lie in the lower half of the substrate UVB \cite{Ruberto}.  

\begin{figure*}
\scalebox{0.79}{\includegraphics{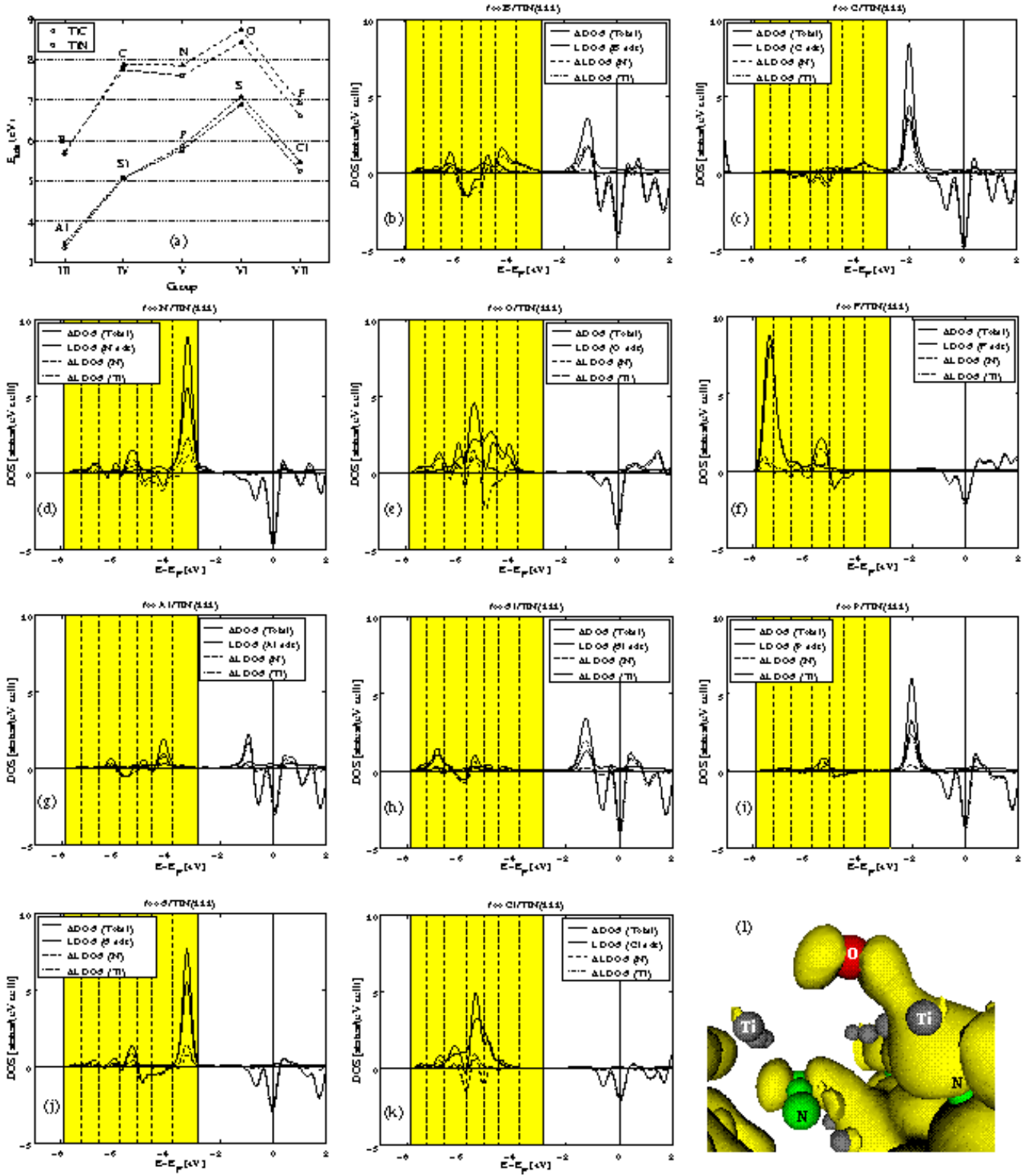}}
\caption{\label{fig:DeltaLDOS_PER2_TiN}Adsorption of the second-period adatoms 
B, C, N, O, and F and of the third-period adatoms Al, Si, P, S, and Cl in fcc site on 
TiX($111$).  
(a) Calculated trends in adsorption energies $E_{\rm ads}$ (excerpt from Ref.\ 
\cite{Vojvodic}).  
(b)--(k) Calculated densities of states (DOS's) for adsorption on TiN($111$).  
Thin line: adatom-projected DOS.  Thick line: difference in total DOS ($\Delta$DOS) 
before and after adsorption for the surface TiN bilayer, including adatom.  
Dashed (dot-dashed) line: $\Delta$DOS projected onto N (Ti) atoms of the surface bilayer.  
Shaded area: energetical location of the substrate UVB.  
Dashed vertical lines: energetical location of the six substrate-UVB peaks.
(l) Contour plot of the KS wavefunction at $E = -6.8$ eV for the fcc O adatom on TiN($111$), 
illustrating a typical bonding adatom--N interaction.}
\end{figure*}

\begin{figure*}
\scalebox{0.96}{\includegraphics{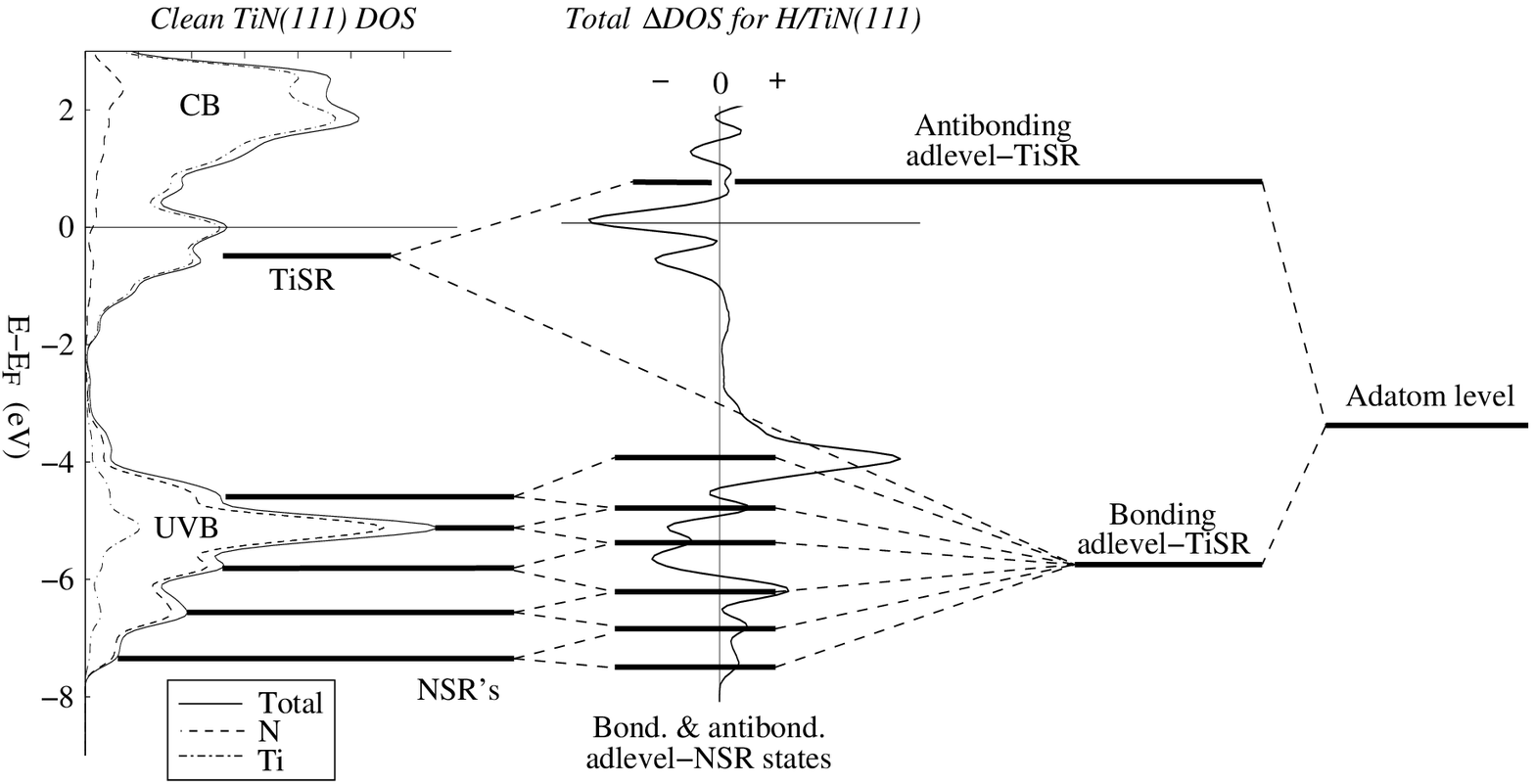}}
\caption{\label{fig:ConcertedCoupling}Schematic diagram of the concerted coupling between 
the adatom state and the TiN($111$) Ti- and N-centered surface resonances (TiSR and NSR's), 
here illustrated for the case of fcc H/TiN($111$).  On the left, the DOS for the clean 
TiN($111$) is reproduced.  In the middle, our calculated $\Delta$DOS for H/TiN($111$) is 
shown, illustrating the quenching and appearance of peaks due to adsorption as 
negative (``$-$'') and positive (``$+$'') peaks, respectively.  Horizontal solid 
lines represent schematically the individual energy levels before and 
after adsorption.}
\end{figure*}

Our new calculations for atomic adsorption on TiN($111$) show that the $\Delta$DOS's 
[Figs.\ \ref{fig:DeltaLDOS_PER2_TiN}(b)--(k)] behave in a similar way on this substrate.  
Figure \ref{fig:ConcertedCoupling} illustrates schematically the concerted-coupling model 
for TiN($111$), using the calculated $\Delta$DOS for fcc H on TiN($111$), which provides the most 
complete illustration of the different types of adatom--substrate couplings.  
As can be seen, the results for adsorption on TiN($111$) show strong similarities to 
those described above for TiC($111$).  
However, a closer comparison reveals differences:  
(i) the antibonding peak above $E_F$ lies at a lower energy than on TiC($111$), $+0.4$ eV; 
(ii) the TiSR quenching is stronger than on TiC($111$) and involves a larger number of TiSR 
electrons; 
(iii) the coupling between the TiSR-modified adlevel and the XSR's occurs at lower 
energies than on TiC($111$).  Points (i) and (ii) can be understood from the 
facts that the TiN($111$) TiSR lies at a lower energy than on TiC($111$) and that this causes a 
larger filling of the TiSR on TiN than on TiC.  Point (iii) is caused by the lower energy 
of the TiN($111$) UVB, compared to that of the TiC($111$) UVB.

In addition, our state-resolved, real-space, analysis of the DOS shows that also 
on TiN($111$) there is a successively increasing dominance of 
adatom--X bonding states [Fig.\ \ref{fig:DeltaLDOS_PER2_TiN}(l)] with decreasing energy.  
Again, however, differences can be detected:  
(i) compared to the adatom--Ti coupling, the adatom--X coupling is generally stronger 
on TiN($111$) than on TiC($111$); 
(ii) while no adatom--Ti coupling can be detected in the TiSR energetical region on TiC($111$), 
such adatom--Ti bonding states are found in the TiSR energetical region on TiN($111$).  
Point (i) can be understood from the higher ionicity of TiN, which results in 
a stronger LDOS on the N atoms, compared to the C LDOS in TiC, and thus a stronger 
adatom--X overlap on TiN($111$) than on TiC($111$).  Point (ii) can be interpreted 
as the presence of a weak coupling, similar to the adatom--UVB coupling, between the 
Ti-localized CB and the antibonding adatom--TiSR states.  Due to the 
CB being partially filled in TiN, this adatom--CB coupling results here 
in filled states, in contrast to TiC, where the CB is empty.  

Our results show thus that the atomic chemisorption on TiN($111$) can be understood and 
analyzed in a way similar to that on TiC($111$).  Both types of interaction of our 
concerted-coupling model (adatom--TiSR and adatom--XSR's) 
appear to be active on both 
substrates, due 
to the similarities between the electronic structures of the two surfaces.  However, some 
differences are detected, 
which can be related to 
the higher ionicity and higher energy of $E_F$ 
in TiN, compared to TiC.  In particular, the results indicate that a weak adatom--CB 
coupling is also present.

\section{Discussion and Conclusions}

Our previous analysis shows that, from the DOS perspective, our concerted-coupling 
model is applicable also to adsorption on TiN($111$).  In the following, we discuss 
how this model can be used to account for the essential features of the calculated 
trends in $E_{\rm ads}$ \cite{Ruberto, Vojvodic}, which varies 
in a characteristic pyramid-like manner for the second-period adatoms B, C, N, O, and F, 
peaking at O, and the third-period adatoms Al, Si, P, S, and Cl, peaking at S, 
on both TiC($111$) and TiN($111$) [Fig.\ \ref{fig:DeltaLDOS_PER2_TiN}(a)].  

Given the strong quenching of the TiSR, a strong contribution to the 
bonding can be expected to come from the 
first component of the concerted coupling, the adatom--TiSR coupling.  
Indeed, the 
adsorption energy for O on the TiSR-deficient TiC($001$) surface is $43\%$ weaker than 
on TiC($111$) \cite{Ruberto}.  The strong bonding contribution of the TiSR arises 
from the fact that the antibonding adatom--TiSR solution lies above $E_F$, thus 
remaining empty for all adatoms.  

Our calculated $\Delta$DOS's [Figs.\ \ref{fig:DeltaLDOS_PER2_TiN}(b--k)] 
show that, within each 
adatom period, the TiSR quenching decreases from group-IV to group-VII 
adatoms, while it increases from group III to group IV.  
These trends could indicate an increase in adatom--TiSR coupling strength from group 
III to group IV and a successive decrease from group IV to group VII.  
These trends agree with the calculated $E_{\rm ads}$ trends for B $\rightarrow$ C, for 
Al $\rightarrow$ Si, for O $\rightarrow$ F, and for S $\rightarrow$ Cl.  
Also, the smaller TiSR quenching in period 3 than in period 2 
indicates a weaker adatom--TiSR coupling for period 3, in agreement with our $E_{\rm ads}$ 
results.  

However, this trend in adatom--TiSR coupling strength disagrees strongly with the 
calculated $E_{\rm ads}$ trends for C $\rightarrow$ N $\rightarrow$ O and for 
Si $\rightarrow$ P $\rightarrow$ S.  On the other hand, our state-resolved DOS's 
(see Sec.\ III.B) show that the second component of the 
concerted coupling, the adatom--XSR couplings, gets successively more active as 
the energetical overlap between the adlevel and the substrate XSR's increases.  
As $Z$ increases within each adatom period, the adlevel energy decreases, thus 
increasing the overlap between the adlevel and the energetical region of the 
XSR's (that is, the lower half of the substrate UVB).  Therefore, the adatom--XSR 
coupling strengths can be expected to increase as $Z$ increases within each adatom 
period, providing an explanation for the overall increasing $E_{\rm ads}$ values 
from group III to group VI.  
In addition, this provides further explanation for the stronger chemisorption 
of the second-period adatoms, compared to the third period, due to the lower free-atom 
energy, within a given group, of the second-period adatoms.  

On the other hand, such a trend is in contrast to the $E_{\rm ads}$ trends for 
O $\rightarrow$ F and S $\rightarrow$ Cl.  
However, F and Cl have ionic bonds, 
as shown by our calculated Bader charges, which yield a charge transfer from the surface Ti 
atoms to the F and Cl adatoms of $0.75$--$0.80$ electrons, and by our calculated electron 
distributions, which show negligible bond charges for F and Cl \cite{Vojvodic}.  
This implies almost fully occupied outer electron shells for the F and Cl adatoms, which can 
therefore be considered to be almost inert toward further chemical interaction, that is, 
toward a coupling with the UVB states.  Indeed, our $\Delta$DOS's show 
that the F and Cl adlevels do not broaden as significantly as the levels of other adatoms 
with the same amount of overlap with the UVB region 
[{\it cf.}, for example, Cl and O on TiN($111$)], indicating a weaker adatom--UVB coupling.  
For all the other fcc adatoms, the electron densities show a clear covalent character for 
the adatom--substrate bond, with the ionic character increasing successively with increasing 
$Z$ within each adatom period.  This is confirmed by our calculated Bader charges for the 
adatoms:  
$-1.07$, $-1.31$, $-1.34$, $-1.20$, and $-0.80$ for the second-period adatoms B, C, N, O, 
and F, respectively (corresponding to $21\%$, $33\%$, $45\%$, $60\%$, and $80\%$ 
of their empty outer electron shells) and $-0.40$, $-0.89$, $-1.08$, $-1.05$, and $-0.74$ 
for the third-period adatoms Al, Si, P, S, and Cl, respectively 
(corresponding to $8\%$, $22\%$, $36\%$, $52\%$, and $74\%$ of their empty outer 
electron shells).  

Thus, the concerted-coupling model, which well describes the calculated DOS 
results, provides also a way for understanding the main features of the 
calculated $E_{\rm ads}$ trends for both TiX($111$) surfaces.  In particular, 
our analysis points toward a greater role for the adatom--XSR couplings in this 
understanding than what has been suggested so far in the literature.

Our results show also a small but interesting difference between TiC and TiN:  
for adatoms with low $Z$ (B and Al) chemisorption is slightly stronger on TiN($111$), 
while for adatoms with higher $Z$ it is stronger on TiC($111$).  
For B and Al, the adlevels lie above the UVB region.  The 
adatom--XSR contributions to their bonding strengths are 
therefore negligible and their chemisorption is dominated by the adatom--TiSR coupling.  
Hence, our $E_{\rm ads}$ results imply that the bonding 
contribution from the adatom--TiSR coupling is somewhat stronger on TiN($111$) than on 
TiC($111$), which is in agreement with our previous observation that the TiSR quenching is 
stronger on TiN than on TiC.  
Now, as stated above, as the adlevel energy shifts down and overlaps with the UVB 
region, the adatom--XSR couplings get 
successively stronger.  However, for the same adlevel 
shift, this overlap is stronger on TiC than on TiN, due to the smaller UVB--CB energy gap of 
TiC.  Therefore, for the same adatom, the 
adatom--XSR contributions to the bonding strength are stronger on TiC than on TiN.  
As $Z$ increases within each adatom period, this effect grows strong enough to 
``neutralize'' the stronger adatom--TiSR coupling for TiN and invert the 
chemisorption preference between TiN and TiC.  

In summary, the application of our concerted-coupling model, originally 
proposed for adsorption on TiC($111$), can be 
generalized to TiN($111$).  However, the 
quantitative differences in electronic structures between 
the two compounds must be taken into account.  
This shows that adsorption on TiC and TiN is more complex and versatile 
than on, for instance, pure metal 
surfaces and that the combination of several different mechanisms must be taken into 
account, giving a potential for 
variations, very useful in applications.

\begin{acknowledgments}

Valuable discussions with \O yvind Borck are gratefully acknowledged, as is financial 
support from the Swedish Foundation for Strategic Research via Materials Consortium \#9 
and ATOMICS and the Swedish Scientific Council, 
as well as allocation of computer time at the UNICC facility (Chalmers) and at 
SNIC (Swedish National Infrastructure for Computing).  

\end{acknowledgments}



\begin{thebibliography}
\bibitem{footnote_TiSR}
These TiSR's have been named surface states in the literature, 
but we prefer to adopt the proper term surface resonances (SR's), since their 
energies do not lie in true bulk band gaps [see Figs.\ \ref{fig:LDOS_surf}(a--b)].
\end{thebibliography}
\end{document}